\documentclass[letterpaper]{article}
\usepackage{verbatim,graphicx,amsmath,amssymb,newalg}

\begin{document}
\title{Alert-BDI: BDI Model with Adaptive Alertness through Situational Awareness}
\author{Manu S Hegde and Sanjay~Singh\thanks {Sanjay Singh is with the Department of Information and Communication Technology, Manipal Institute of Technology, Manipal University, Manipal-576104, INDIA \hspace{5cm}E-mail: sanjay.singh@manipal.edu}}

\maketitle

\begin{abstract}
In this paper, we address the problems faced by  a group of agents that possess situational awareness, but lack a security mechanism, by the introduction of a adaptive risk management system. The Belief-Desire-Intention (BDI) architecture lacks a framework that would facilitate an adaptive risk management system that uses the situational awareness of the agents. We extend the BDI architecture with the concept of adaptive alertness. Agents can modify their level of alertness by monitoring the risks faced by them and by their peers. Alert-BDI enables the agents to detect and assess the risks faced by them in an efficient manner, thereby increasing operational efficiency and resistance against attacks.
\end{abstract}


\section{Introduction}
Autonomous multi-agent systems represent a new approach for the analysis, design, and implementation of complex software systems \cite{A_Roadmap}. Autonomy, robustness and distributiveness are the major features of multi-agent systems. These features makes multi-agent systems an ideal choice for the implementation of various complex systems. The agent-based systems may be used in the conceptualization and implementation of many types of software systems. Multi-agent systems are being used in a variety of fields ranging from network security to search-and-rescue systems.

\par 
In many real world scenarios, multi-agent systems needs to operate without global knowledge and global communications, that is, the agents operate in the absence of situational awareness. Global knowledge refers to information that is available to all agents in the system at any instant of time. However information that an agent gains by perceiving its local environment remains localized unless it is communicated to other agents. Also, in an insecure environment, an agent would not know whether to believe or not the information it has retrieved from a peer agent. Global communications refers to the ability of an agent to communicate with any peer agents in the system at any instant. However communications are costly with respect to time and energy. Local information gained by an agent may have little value to an agent operating at a different location.

\par 
Situational awareness is key to operate securely in an insecure environment. Awareness may be gained by agents in a multi-agent system through inter-agent communication as described in \cite{Aware}. Situational awareness leads to enhanced operational efficiency of the system. The awareness mechanism enables the agents to share data by sending and retrieving beliefs. Hegde and Singh \cite{Aware}, has proposed a mechanism that allows agents to test the relevancy of beliefs, so that only relevant beliefs would be shared, thereby reducing communications and hence reducing the communication overheads. Situational awareness empowers individual agents with the ability to ascertain the credibility of the information received and the reliability of the agent that supplied it.

\par 
In many real world scenarios, multi-agent systems are required to incorporate security into their operations, for the efficient and successful achievement of their goals. Incorporating security into agent operations is a drain on the agent's resources. However it is vital to the agent's success, in an insecure environment. To lessen the impact of security operations on the agent's resources, the agent may employ a security mechanism that adapts itself according to the threats perceived by the agent. Situational awareness may be used to analyze perceived information to detect and prioritize threats. The agents have to incorporate a risk management system that analyzes information acquired by the situational awareness system, detect threats and classify them, and calculate the risk posed to the agent by these threats. This enables the agent's security system to change the intensity of its operations in view of the risks faced by the agent.

\par 
To enable the use of autonomous agents in the modeling and design of complex systems, the agents should have the ability to perceive the environment and directs its activity towards achieving its objectives. Many practical reasoning agent models have been proposed in recent years. The BDI model is one such reasoning agent model that tries to mimic human reasoning using the concepts of beliefs, desires and intentions. A BDI architecture addresses how these components are represented, updated, and processed to determine the agent's actions \cite{Sardina}. Many agent programming languages and development platforms use the BDI architecture, such as PRS \cite{PRS}, dMARS \cite{dMARS}, AgentSpeak \cite{AgentSpeak}, Jason \cite{Jason}, JADEX \cite{Jadex}, GOAL \cite{Goal}, Jack \cite{Jack} and JAM \cite{Jam}. However the absence of an adaptive security mechanism based on situational awareness limits the possibilities of designing efficient multi-agent BDI systems that need to operate in an insecure environment.

\begin{figure}[ht]
	\centering
		\includegraphics[width=9cm]{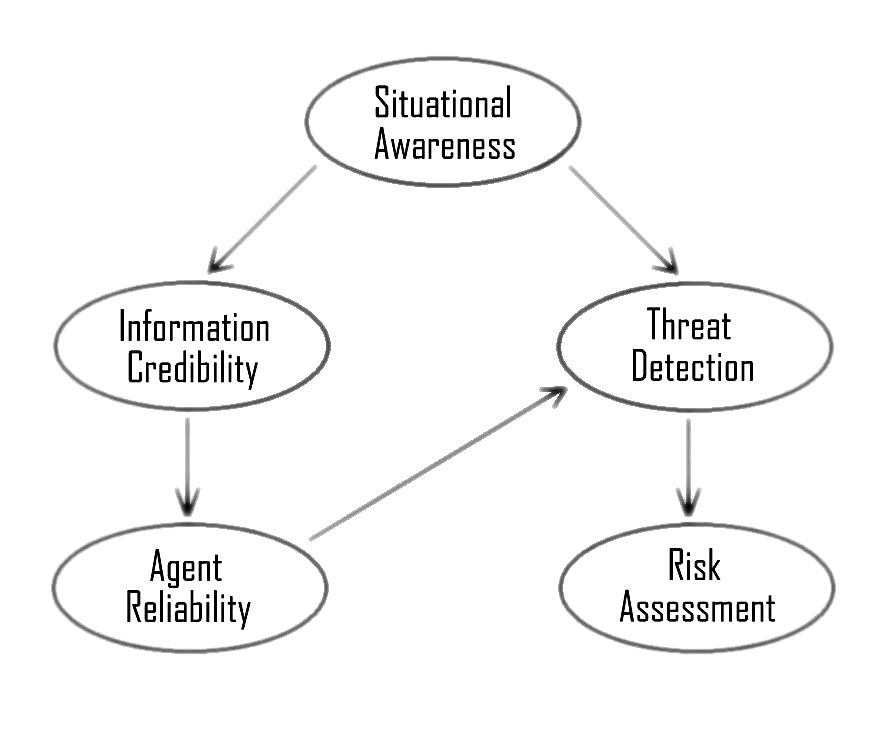}
		\caption{Working of Alert-BDI extension}
	\label{fig:A1}
\end{figure}

In this paper, we propose an extension to the existing BDI model \cite{bdi}, that helps the agents analyze the data obtained through situational awareness, and detect threats to evaluate the risk posed to them by various threats in their vicinity thereby leading to enhanced operational efficiency as represented in Fig. 1. This extension will enable the agents:
\begin{itemize}
\item To analyze the situational awareness data and identify threats based on agent behavior. Agent behavior may be the truthfulness or responsiveness of the agents. Threats may be assigned with priorities based on the risk posed by them.
\item To assess the risks faced, by analyzing the perceived threats and risk assessments of peer agents. The risk assessments of peer agents are considered based on agent reputations.
\end{itemize}
Rest of the paper is organized as follows. Section II briefs the related work in the area. Section III briefly describes about the existing BDI model and glowworm swarm optimization. Section IV explains the proposed Alert-BDI and finally section V concludes this paper.
\section{Related Work}
The BDI architecture provides an abstract reasoning system that enables the modeling and implementation of complex systems as an interaction of intelligent agents using the notion of beliefs, desires and intentions. However, due to the lack of a complete specification of the BDI architecture, many systems have been developed that claim to conform to the BDI model, but differ from it in key aspects. Inverno et al \cite{A_Formal} proposed an abstract formal model, dMARS, for the BDI architecture. It defines the key data structures and operations that manipulate these structures, to enable the modeling of intelligent agents based on the BDI model.

\par 
The belief-desire-intention (BDI) model provides an architecture to enable the creation of intelligent agents, that can be used to analyze, design and implement complex systems. Many extensions have been proposed in recent years, to extend the functionality of the BDI model. These extensions enhance the usefulness of the basic BDI model. Panzarasa, Norman and Jennings \cite{MSBF} introduced the concept of social mental shaping, which extends the BDI model to account for how common social phenomena can influence an agent's problem-solving behavior. It investigates the impact of social nature of agents upon their individual mental states. The concept of cooperation among agents is vital to detect threats in the neighborhood of the agent, in a dynamic environment. But this system does not explore how the cooperation may be extended to bring about awareness and security for the agents.

\par 
The concept of situational awareness has been introduced Hegde and Singh \cite{Aware}, which enables agents to share beliefs with their peers in an efficient manner. The agents share data with a select group of its neighbors, formed using Glow-worm swarm optimization to optimize the belief sharing mechanism. It proposes belief agreement and agent reputation algorithms, that use Fleiss' Kappa \cite{kappa} to analyses the veracity of data shared by the peer agents. The agents may use information gained about the reliability of an agent and its general behavior to analyses whether an agent is a threat or not. Though this system, analyzes techniques to detect false communications by agents, it does not make any attempt to use them to detect threats or analyze the risks faced by individual agents.

\par Cooperative multi-agent systems jointly solve tasks through their interaction to maximize utility. Due to the interactions among the agents, multi-agent problem complexity is directly proportional to the number of agents and agent behavioral complexity \cite{Panait}. This requires the the multi-agent systems to optimize their interactions, to enable the agents to operate in an efficient manner. Lower interactions may result in the agents missing out on vital information. Higher interaction among the agents increases the complexity of the system, leading to reduced system efficiency. To solve this problem, we propose the use of a swarm optimization technique to optimize the risk assessment procedure.

\par 
The proposed system uses the concept of situational awareness proposed in \cite{Aware}, to develop a threat detection and risk assessment mechanism. This would enable agents to collaborate with their peers to detect anomalous agent behavior. Moreover, the risk assessment mechanism allows the agents to perceive the risk posed to them, due to the threats detected. The risk assessment technique is optimized by the use of glowworm swarm optimization \cite{GSO_mommf}. This is crucial to ensure scalability of the system.

\section{Theoretical Background}

This section describes the BDI model and Glowworm Swarm Optimization.

\subsection{The BDI Model}
The belief–-desire–-intention model is a model developed for providing agents with a reasoning ability. The BDI model implements the principal aspects of Michael Bratman's theory of human practical reasoning \cite{Wiki_BDI}. The key components of the BDI model include beliefs, desires and intentions. Beliefs represent the information available to an agent. The information may include its observations of the world or communicational data from other agents. Desires represent the objectives that the agent would like to accomplish. A desire to which an agent commits itself, forms its intention. Intentions are courses of action that an agent has committed to carry out \cite{Learning_in}. Current intentions were influenced by the agent's beliefs and past intentions. Moreover current intentions, constrain the adoption of new intentions.

\subsection{Glowworm Swarm Optimization}
Glowworm Swarm Optimization (GSO) is a new swarm intelligence based algorithm for optimizing multi-modal functions. In GSO, the individuals in the swarm use a dynamic decision domain to effectively locate the local maxima of the function. Each individual in the swarm uses the decision domain to choose its neighbors and decides its direction of movement by the strength of the signal picked up from them. These movements that are based only on local information enable the multi-agent swarm to partition into disjoint subgroups that converge to multiple optima of a given multimodal function \cite{GSO_mommf}.

\par 
The agents are initially placed randomly in the optimization problem's search space. Also, all agents are assigned the same fitness values. Each execution cycle of the algorithm updates individual fitness values, location and decision domain attributes. The details of the GSO algorithm can be found in \cite{GSO_ifga}. The GSO algorithm has been used here to optimize the risk assessment process.

\section{Proposed Method}
\subsection{Overview of Alert-BDI}
Alert-BDI is based on the dMARS specification \cite{A_Formal} and extends the basic BDI architecture by enabling the agents to use situational awareness to achieve adaptive alertness. Situational awareness can be gained through exchange of beliefs, belief relevancy, belief agreement and agent reputation \cite{Aware}. Alertness is the state of being aware of the occurrences of events, and processing the events to detect any unusual and potentially dangerous circumstances. It combines information gathering and analysis of the gathered information to detect threats. Information gathering is the awareness gained by perceiving the environment and retrieving data from its peers. Threat detection is based on agent behavior, that is its interactions with other agents.

\par Malicious agents are not a threat solely to the agents interacting with it but also to other agents in the swarm. Thus, there is an inherent need for agents to collaborate and learn from each other's experience in dealing with malicious agents. Jean et al \cite{Boosting} proposed a monitoring system is that allows agents to learn and classify agents, not only based on their own observations, but also based on collaboration with other agents in the network. To model the malicious intent of agents, we propose an anomaly detection model, that takes into consideration both the responsiveness of an agent and its truthfulness. Responsiveness is a useful metric, since it helps gauge the willingness of an agent to cooperate with its peers. Also, the truthfulness of an agent may be determined to a greater extent if the agent is responsive. Moreover, in a volatile real-world environment, the validity of truth may change rapidly, and hence the anomaly detection model should take this into consideration, while classifying the agents.

\subsection{Structural Specification of Alert-BDI}
Traditional security techniques can protect agents from certain kinds of attacks. However the autonomy and distributiveness of multi-agent systems require the agents to communicate and cooperate among themselves, to tackle the threats faced by them. T.Y Li and K.Y Lam \cite{Detecting} gave an anomaly detection model for detecting malicious agents, that analyzes mobile agent's activity by measuring its movement pattern and residence time on hosts. Observation shows that majority of the agents follow some regular behavior patterns. Hence, anomaly detection may be used to detect agents behaving in an abnormal manner.

\begin{figure}[ht]
	\centering
		\includegraphics[width=9cm]{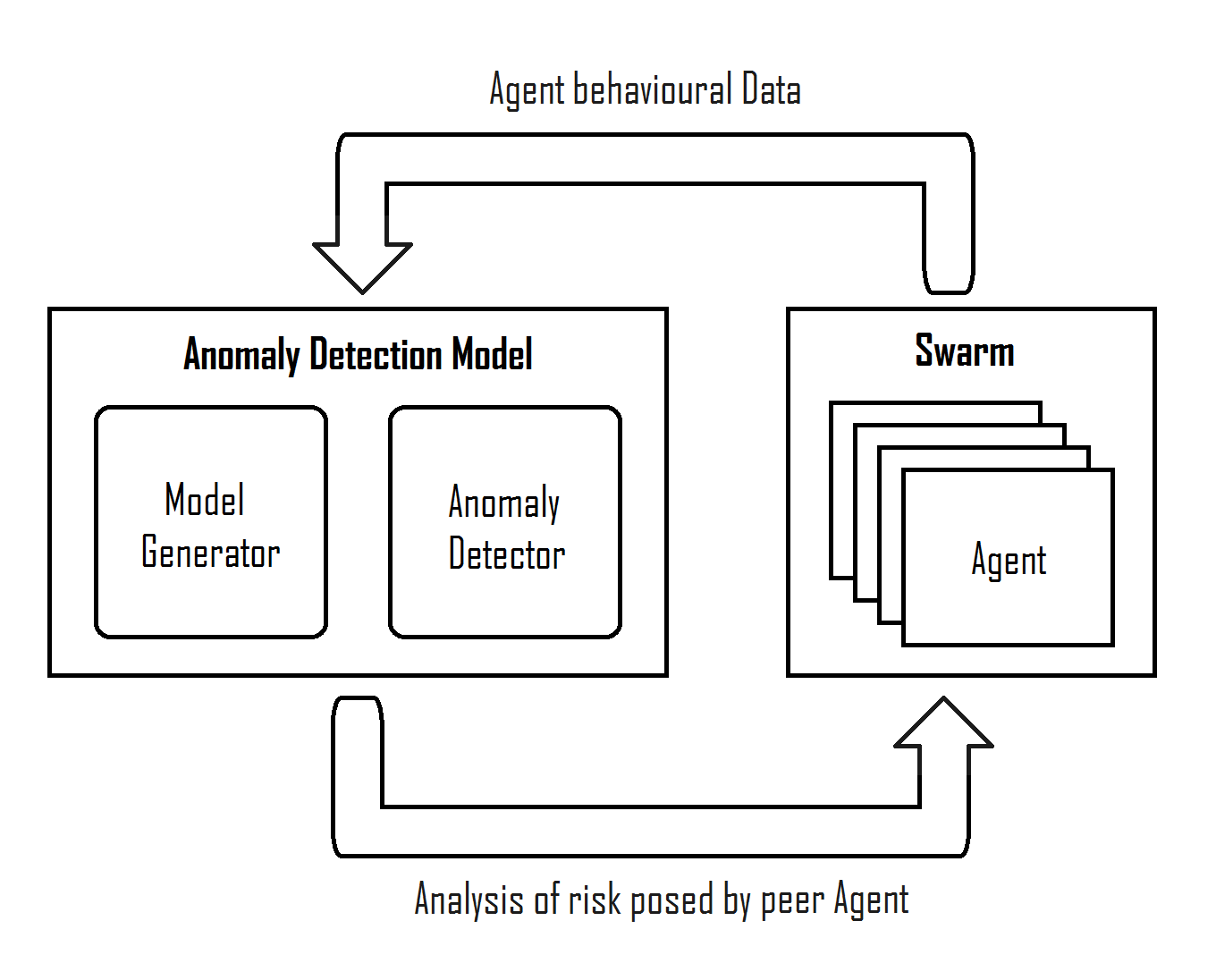}
		\caption{Anomaly Detection Model}
	\label{fig:A2}
\end{figure}

The proposed anomaly detection model processes the information gathered by situationally aware agents. This information may be processed to analyze the perceptions of risk as perceived by peer agents. The information includes truthfulness and responsiveness of agents. Each individual agent scans its local environment to identify peer agents and their behavior. The agent may communicate with its peers over time. These communications are processed to analyze the truthfulness and responsiveness of agents. 
\par 
The anomaly detection model includes a Model Generator and an Anomaly Detector as shown in Fig 2. The model generator receives behavioral data from peer agents and generates agent models periodically. The anomaly detector receives processed data from the model generator, its own perceptions and classifies the threat posed due to other agents.

\par 
Each agent may access agent behavioral data from its peers within its communication range. With increase in the size of the swarm, the number of neighbors within the communication range increases and affects the efficiency of risk assessment. Hence, there is a need to optimize the way agent behavioral data is accessed, so that the efficiency of risk assessment process is enhanced. 
\par
The agents may use glowworm swarm optimization algorithm to identify agents, whose perceived risks may be considered to model the risk faced by it. Also, agent behavior may change with changing situations, and hence the risk assessment algorithm should consider volatile agent behavior while assessing the risk posed to an agent.

\begin{figure}[ht]
	\centering
		\includegraphics[width=9cm]{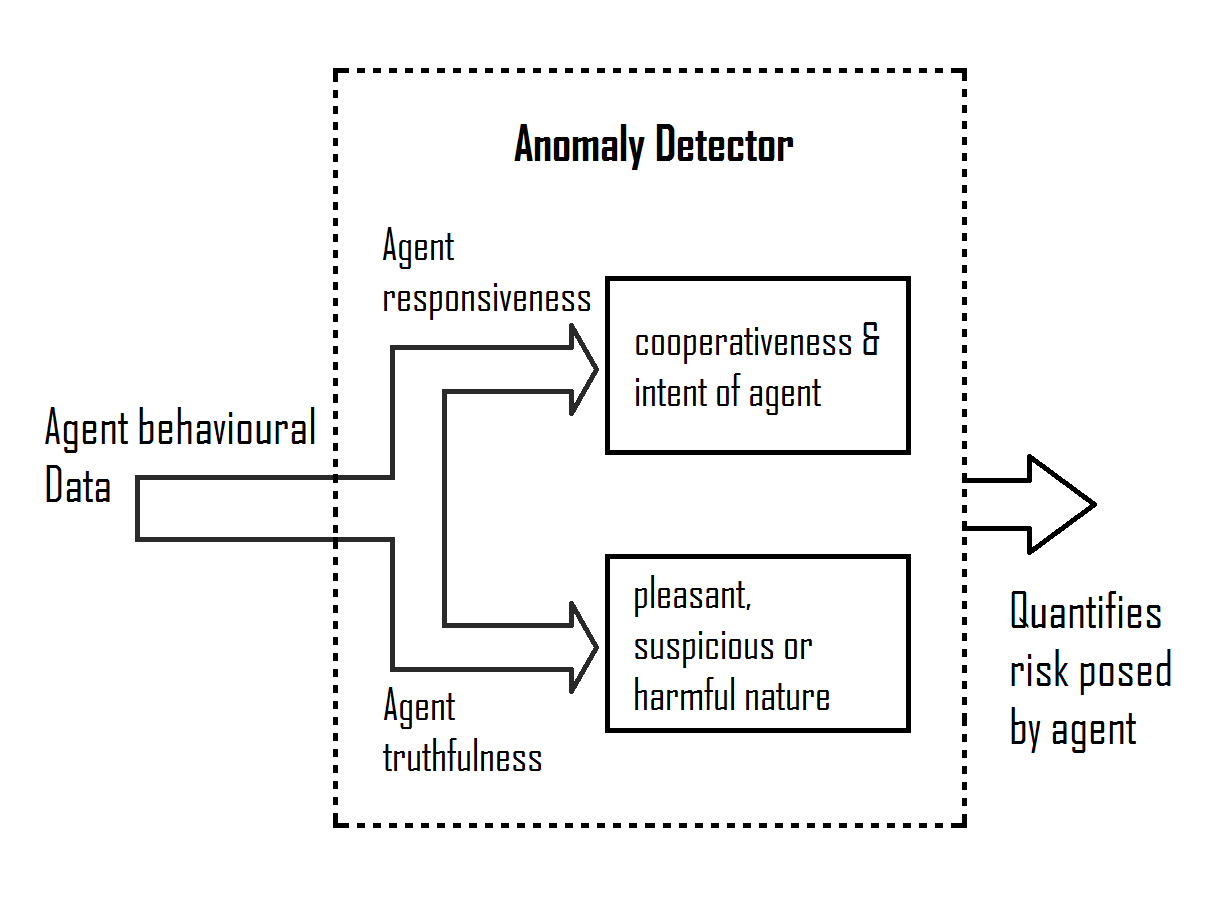}
		\caption{Anomaly Detector}
	\label{fig:A3}
\end{figure}

Algorithm to \textit{merge agent behavioral data}

\begin{algorithm}{mergeBehaviourData}{}
\textit{/* {precondition}: received agent behavioral data} \\
\textit{from a selected group of agents */} \\
\textit{/* postcondition: assigned behavioral data to} \\
\textit{individual agents */} \\
n \= \#(\textit{Neighbor Agents})\\
\begin{FOR}{i \= 1 \TO n}
comAgents \= \textit{ComDomain(Agent i)}\\
\end{FOR}\\
\begin{FOR}{i \= 1 \TO \#(comAgents)}
\begin{FOR}{j \= 1 \TO n}
data \= \textit{Repute(Agent i)} * agentData[i][j]\\
behavior[j] \= behavior[j] + data
\end{FOR}
\end{FOR}
\end{algorithm}

\subsection{Anomaly Detection}
The Anomaly Detection Model (ADM) received agent behavioral data from individuals in the swarm, and provides them with an estimation of the risks they are subjected to. The ADM consists of the Model Generator and an Anomaly Detector. The Model Generator merges agent behavioral data received from peer agents by assigning weights based on agent reputations. 
\par
The Anomaly Detector uses the behavioral data synthesized by the Model Generator, and estimates the threat category of the agent as represented in Fig 3. The agent behavioral data used by the Anomaly Detector to detect anomalous behavior includes responsiveness and truthfulness of agents. The responsiveness factor refers to the willingness of an agent to respond to communications by a peer agent. An agent that is non responsive is at best a non-cooperative agent. But it may be a malicious agent as well, and peer agents do not have any technique to verify its truthfulness. An agent is considered cooperative if it is both responsive and truthful.

\par 
The \textit{estimateAgentThreat()} algorithm estimates the level of threat posed by an agent by evaluating its responsiveness and truthfulness. It defines four classes of threat levels, namely Cooperative, Suspicious, Malicious and Noxious. The threat level assigned to an agent is volatile, since a change in agent behavior will affect its threat level. Cooperative agents are friendly and pleasant in their interaction with their peers. An agent is labeled suspicious, if it is unresponsive despite being truthful, since its unresponsiveness makes it difficult for its peers to detect a change in its truthfulness. An agent is said to be malicious if the agent has been untruthful in the past and its intentions are not clear, due to its unresponsiveness. A noxious agent is clearly harmful, since it disseminating false information to its peers.\\

\par Algorithm to \textit{estimate the threat level of an agent}

\begin{algorithm}{estimateAgentThreat}{}
\textit{/* {precondition}: received merged responsiveness} \\
\textit{and truthfulness data of agents */} \\
\textit{/* postcondition: assigned threat level to} \\
\textit{individual agents */} \\
n \= \#(\textit{Neighbor Agents})\\
\begin{FOR}{i \= 1 \TO n}
\begin{IF}{responsive[i] > respondThreshold}\\
\begin{IF}{truthful[i] > truthThreshold}\\
\text{classify agent as Cooperative}
\ELSE\\
\text{classify agent as Noxious}
\end{IF}
\ELSE\\
\begin{IF}{truthful[i] < truthThreshold}\\
\text{classify agent as Malicious}
\ELSE\\
\text{classify agent as Suspicious}
\end{IF}
\end{IF}
\end{FOR}\\
\end{algorithm}

\subsection{Risk Assessment}
Alert-BDI uses the GSO algorithm, to optimize the way agent behavioral data is retrieved from peer agents, so that the efficiency of risk assessment process is enhanced. The GSO algorithm is used to select a communication domain, and optimize communication domain members based on agent truthfulness. The agents in the system, will identify a group of agents among their neighbors, who will be contacted to retrieve agent behavioral data, to ensure efficiency. 
\par
The fitness of each agent will be based on its probability of being truthful in its communications. If an agent responds to communication from its neighbors and its peers validate its response as being truthful, it is more likely to be included in the communication groups of other agents. The fitness updation function that decides the value of agent $i$ is given by \cite{GSO_mommf}:

\begin{equation}
G_i(t)=(1-\rho)G_i(t-1)+\gamma J_i(t)
\end{equation}
where, $G_i(t)$ is new amount of luciferin, $G_i(t-1)$ is former amount of luciferin and $J_i(t)$ is fitness of agent $i$ in iteration t of the algorithm and $\rho$ and $\gamma$ are constants for modeling luciferin gradual drop and the effect of fitness on luciferin. $J_i(t)$ is the degree of truthfulness shown by agent $i$ to its neighbors.

For every agent $i$, the probability of moving into neighboring agent $j$'s communication domain is defined as,
\begin{equation}
P_{ij}(t)=\frac{G_{j}(t)-G_{i}(t)}{\sum\sb{k\in N_i(t)}G_k(t)-G_i(t)}
\end{equation}

where $N_i(t)$ is defined as,
\begin{equation}
N_i(t)=\{j:d_{ij}<r\sb{d}\sp{i}(t);G_i(t)<G_j(t)\}\ \mbox{and}\ j\in N_i(t)
\end{equation}
where $d_{ij}$ is the distance between agents $i$ and $j$.

\par Each agent $i$ maintains communications with a select group of its neighbors; the communication domain $N_i(t)$ of agent $i$ consists of those agents that have a relatively higher fitness value and that are located within a dynamic decision domain whose range, $r\sb{d}\sp{i}(t)$, is bounded by a circular sensor range\\ $(0 < r\sb{d}\sp{i}(t)\leq r_s)$. The decision domain range for each agent $i$ is updated using the expression,
\begin{equation}
r\sb{d}\sp{i}(t+1)=min\{r_s,max\{0,r\sb{d}\sp{i}(t)+\beta(n_t-|N_i(t)|)\}\}
\end{equation}
where, $\beta$ is a constant parameter and $n_t$ is a parameter to control the number of agents in the decision domain.\\

Algorithm to determine Communication domain

\begin{algorithm}{comDomain}{\textit{agent j}}
\textit{/* {precondition}: agent $j$ is in the neighborhood of } \\
\textit{the agent */} \\
\textit{/* postcondition: agent $j$ is included or excluded from } \\
\textit{the communication domain */} \\
\textit{agent i} \= \textit{agent executing comDomain()}\\
\textit{/* Luciferin updating phase */}\\
G_i(t)=(1-\rho)G_i(t-1)+\gamma J_i(t)\\
\textit{/* Communication Domain updating phase */}\\
\textit{/* Determining probability of agent $j$ being included in}\\
\textit{the Communication domain */} \\
P_{ij}(t) \= \frac{G_{j}(t) - G_{i}(t)}{\sum\sb{k\in N_i(t)}G_k(t) - G_i(t)}\\
N_i(t) \= j:d_{ij}<r\sb{d}\sp{i}(t);G_i(t)<G_j(t)\\
maxAgents \= s\\
\begin{WHILE}{\#(N_i(t)) > s}\\
\begin{FOR}{k \= 1 \TO \#(N_i(t))}\\
rAgent \= \textit{agent k with leastVal($P_{ij}(t)$)}
\end{FOR}\\
\textit{remove agent rAgent from $N_i(t)$}
\end{WHILE}\\
\textit{/* updating range of Communication Domain */}\\
r\sb{d}\sp{i}(t+1) \= min\{r_s,max\{0,r\sb{d}\sp{i}(t)+\beta(n_t-|N_i(t)|)\}\}
\end{algorithm}

\par The communication domain of an agent is the subset of the agent's neighbors from whom the agent would retrieve agent behavioral data. The \textit{comDomain()} algorithm is based on the GSO algorithm. It consists of three phases, namely luciferin updation phase, communication domain updation phase and the communication domain range updation phase. The luciferin updation phase is concerned with updating the fitness of the agent based on past and current fitness values.
\par
In the communication domain updation phase the agent first determines the probability of agent $j$ being included in its communication domain. This is calculated based on the luciferin values of agents $i$ and $j$ and that of the agents in $i$'s communication domain. Then the communication domain is updated based on the communication domain range of agent $i$ and the luciferin values of agents $i$ and $j$. If the number of agents in its communication domain exceeds a predefined limit, it discards a member agent that has the lowest probability of inclusion.
\par
In the communication domain range updation phase, the agent updates the range of its communication domain based on its sensor range, past value of its communication domain range and the current size of its communication domain.

\section{Conclusion}
In the basic BDI model, agents can exchange beliefs by sending messages to peer agents. However, the absence of a mechanism that enables a multi-agent system to detect threats and assess risks in an efficient and cooperative manner, makes them vulnerable to attacks.

\par 
In this paper, we propose an extension to the basic BDI model that would enable agents to achieve adaptive alertness using situational awareness. Achieving adaptive alertness that adapts to changing conditions in the environment, enables the agents to employ a security mechanism that can vary to suit the risk faced. Such a variable security mechanism saves the agent's resources for the fulfillment of their goals.

\par 
Adaptive alertness achieves the task of threat detection and risk assessment by gaining information through situational awareness among the agents. The agents may communicate with their peers to evaluate agent's behavior in a collaborated effort. Moreover, the use of glowworm swarm optimization to optimize the risk assessment process enhances the scalability of the overall system.

\bibliographystyle{IEEEtran}
\bibliography{myref}

\begin{thebibliography}{10}
\providecommand{\url}[1]{#1}
\csname url@samestyle\endcsname
\providecommand{\newblock}{\relax}
\providecommand{\bibinfo}[2]{#2}
\providecommand{\BIBentrySTDinterwordspacing}{\spaceskip=0pt\relax}
\providecommand{\BIBentryALTinterwordstretchfactor}{4}
\providecommand{\BIBentryALTinterwordspacing}{\spaceskip=\fontdimen2\font plus
\BIBentryALTinterwordstretchfactor\fontdimen3\font minus
  \fontdimen4\font\relax}
\providecommand{\BIBforeignlanguage}[2]{{%
\expandafter\ifx\csname l@#1\endcsname\relax
\typeout{** WARNING: IEEEtran.bst: No hyphenation pattern has been}%
\typeout{** loaded for the language `#1'. Using the pattern for}%
\typeout{** the default language instead.}%
\else
\language=\csname l@#1\endcsname
\fi
#2}}
\providecommand{\BIBdecl}{\relax}
\BIBdecl

\bibitem{A_Roadmap}
\BIBentryALTinterwordspacing
N.~R. Jennings, K.~Sycara, and M.~Wooldridge, ``A roadmap of agent research and
  development,'' \emph{Autonomous Agents and Multi-Agent Systems}, vol.~1,
  no.~1, pp. 7--38, Jan. 1998. [Online]. Available:
  \url{http://dx.doi.org/10.1023/A:1010090405266}
\BIBentrySTDinterwordspacing

\bibitem{Aware}
M.~S. Hegde and S.~Singh, ``Aware-{BDI}: An extension of {BDI} model
  incorporating situational awareness,'' in \emph{Third International
  Conference on Communication Systems and Network Technologies 2013 (CSNT
  2013)}, Gwalior, India, April 2013, pp. 100--104.

\bibitem{Sardina}
\BIBentryALTinterwordspacing
S.~Sardina and L.~Padgham, ``A bdi agent programming language with failure
  handling, declarative goals, and planning,'' \emph{Autonomous Agents and
  Multi-Agent Systems}, vol.~23, no.~1, pp. 18--70, Jul. 2011. [Online].
  Available: \url{http://dx.doi.org/10.1007/s10458-010-9130-9}
\BIBentrySTDinterwordspacing

\bibitem{PRS}
F.~Ingrand, M.~Georgeff, and A.~Rao, ``An architecture for real-time reasoning
  and system control,'' \emph{IEEE Expert}, vol.~7, no.~6, pp. 34--44, Dec.

\bibitem{dMARS}
\BIBentryALTinterwordspacing
M.~D'Inverno, M.~Luck, M.~Georgeff, D.~Kinny, and M.~Wooldridge, ``The dmars
  architecture: A specification of the distributed multi-agent reasoning
  system,'' \emph{Autonomous Agents and Multi-Agent Systems}, vol.~9, no. 1-2,
  pp. 5--53, Jul. 2004. [Online]. Available:
  \url{http://dx.doi.org/10.1023/B:AGNT.0000019688.11109.19}
\BIBentrySTDinterwordspacing

\bibitem{AgentSpeak}
D.~Silva and J.~Gluz, ``Agentspeak(pl): A new programming language for bdi
  agents with integrated bayesian network model,'' in \emph{Information Science
  and Applications (ICISA), 2011 International Conference on}, April, pp. 1--7.

\bibitem{Jason}
\BIBentryALTinterwordspacing
S.~Vester, N.~Boss, A.~Jensen, and J.~Villadsen,
  ``\BIBforeignlanguage{English}{Improving multi-agent systems using jason},''
  \emph{\BIBforeignlanguage{English}{Annals of Mathematics and Artificial
  Intelligence}}, vol.~61, pp. 297--307, 2011. [Online]. Available:
  \url{http://dx.doi.org/10.1007/s10472-011-9225-2}
\BIBentrySTDinterwordspacing

\bibitem{Jadex}
\BIBentryALTinterwordspacing
L.~Braubach, A.~Pokahr, and W.~Lamersdorf, ``Jadex: A bdi-agent system
  combining middleware and reasoning,'' in \emph{Software Agent-Based
  Applications, Platforms and Development Kits}, ser. Whitestein Series in
  Software Agent Technologies, R.~Unland, M.~Calisti, and M.~Klusch, Eds.\hskip
  1em plus 0.5em minus 0.4em\relax Birkhäuser Basel, 2005, pp. 143--168.
  [Online]. Available: \url{http://dx.doi.org/10.1007/3-7643-7348-2\_7}
\BIBentrySTDinterwordspacing

\bibitem{Goal}
\BIBentryALTinterwordspacing
K.~Hindriks, ``\BIBforeignlanguage{English}{A verification logic for goal
  agents},'' in \emph{\BIBforeignlanguage{English}{Specification and
  Verification of Multi-agent Systems}}, M.~Dastani, K.~V. Hindriks, and
  J.-J.~C. Meyer, Eds.\hskip 1em plus 0.5em minus 0.4em\relax Springer US,
  2010, pp. 225--254. [Online]. Available:
  \url{http://dx.doi.org/10.1007/978--1--4419--6984--2\_8}
\BIBentrySTDinterwordspacing

\bibitem{Jack}
\BIBentryALTinterwordspacing
M.~Winikoff, ``Jack™ intelligent agents: An industrial strength platform,'' in
  \emph{Multi-Agent Programming}, ser. Multiagent Systems, Artificial
  Societies, and Simulated Organizations, R.~Bordini, M.~Dastani, J.~Dix, and
  A.~Fallah~Seghrouchni, Eds.\hskip 1em plus 0.5em minus 0.4em\relax Springer
  US, 2005, vol.~15, pp. 175--193. [Online]. Available:
  \url{http://dx.doi.org/10.1007/0--387--26350--0\_7}
\BIBentrySTDinterwordspacing

\bibitem{Jam}
\BIBentryALTinterwordspacing
M.~J. Huber, ``Jam: a bdi-theoretic mobile agent architecture,'' in
  \emph{Proceedings of the third annual conference on Autonomous Agents}, ser.
  AGENTS '99.\hskip 1em plus 0.5em minus 0.4em\relax New York, NY, USA: ACM,
  1999, pp. 236--243. [Online]. Available:
  \url{http://doi.acm.org/10.1145/301136.301202}
\BIBentrySTDinterwordspacing

\bibitem{bdi}
M.~Bratman, \emph{Intention, Plans, and Practical Reason}.\hskip 1em plus 0.5em
  minus 0.4em\relax Center for the {S}tudy of {L}anguage and {I}nformation, May
  1999.

\bibitem{A_Formal}
\BIBentryALTinterwordspacing
M.~d'Inverno, D.~Kinny, M.~Luck, and M.~Wooldridge, ``A formal specification of
  dmars,'' in \emph{Proceedings of the 4th International Workshop on
  Intelligent Agents IV, Agent Theories, Architectures, and Languages}, ser.
  ATAL '97.\hskip 1em plus 0.5em minus 0.4em\relax London, UK, UK:
  Springer-Verlag, 1998, pp. 155--176. [Online]. Available:
  \url{http://dl.acm.org/citation.cfm?id=648204.749438}
\BIBentrySTDinterwordspacing

\bibitem{MSBF}
P.~Panzarasa, T.~J. Norman, and N.~R. Jennings, ``Modeling sociality in the bdi
  framework,'' in \emph{In Proceedings of First Asia-Pacific Conference on
  Intelligent Agent Technology (IAT'99}, 1999, pp. 202--206.

\bibitem{kappa}
\BIBentryALTinterwordspacing
Wikipedia. (2013) Fleiss' kappa. [Online]. Available:
  \url{http://en.wikipedia.org/wiki/Fleiss\%27\_kappa}
\BIBentrySTDinterwordspacing

\bibitem{Panait}
\BIBentryALTinterwordspacing
L.~Panait and S.~Luke, ``Cooperative multi-agent learning: The state of the
  art,'' \emph{Autonomous Agents and Multi-Agent Systems}, vol.~11, no.~3, pp.
  387--434, Nov. 2005. [Online]. Available:
  \url{http://dx.doi.org/10.1007/s10458-005-2631-2}
\BIBentrySTDinterwordspacing

\bibitem{GSO_mommf}
\BIBentryALTinterwordspacing
K.~N. Krishnanand and D.~Ghose, ``Glowworm swarm optimisation a new method for
  optimising multimodal functions,'' \emph{Int. J. Comput. Intell. Stud.},
  vol.~1, no.~1, pp. 93--119, May 2009. [Online]. Available:
  \url{http://dx.doi.org/10.1504/IJCISTUDIES.2009.025340}
\BIBentrySTDinterwordspacing

\bibitem{Wiki_BDI}
``Belief-–desire–-intention software model,'' [Available Online]
  http://en.wikipedia.org/wiki/Belief-desire-intention\_software\_model, 2012.

\bibitem{Learning_in}
A.~Guerra-Hern\'{a}ndez, A.~El~Fallah-Seghrouchni, and H.~Soldano, ``{Learning
  in BDI multi-agent systems},'' in \emph{Proceedings of the 4th international
  conference on Computational Logic in Multi-Agent Systems}, ser. CLIMA
  IV'04.\hskip 1em plus 0.5em minus 0.4em\relax Berlin, Heidelberg:
  Springer-Verlag, 2004, pp. 218--233.

\bibitem{GSO_ifga}
\BIBentryALTinterwordspacing
B.~Iranpour and M.~Meybodi, ``An improved fuzzy based glowworm algorithm,''
  \emph{International Journal of Engineering and Technology}, vol.~2, no.~5,
  pp. 900--905, May 2012. [Online]. Available:
  \url{http://iet-journals.org/archive/2012/may\_vol\_2\_no\_5/3891133114277.pdf}
\BIBentrySTDinterwordspacing

\bibitem{Boosting}
\BIBentryALTinterwordspacing
E.~Jean, Y.~Jiao, A.~R. Hurson, and T.~E. Potok, ``Boosting-based distributed
  and adaptive security-monitoring through agent collaboration,'' in
  \emph{Proceedings of the 2007 IEEE/WIC/ACM International Conferences on Web
  Intelligence and Intelligent Agent Technology - Workshops}, ser. WI-IATW
  '07.\hskip 1em plus 0.5em minus 0.4em\relax Washington, DC, USA: IEEE
  Computer Society, 2007, pp. 516--520. [Online]. Available:
  \url{http://dl.acm.org/citation.cfm?id=1339264.1339633}
\BIBentrySTDinterwordspacing

\bibitem{Detecting}
\BIBentryALTinterwordspacing
T.-Y. Li and K.-Y. Lam, ``Detecting anomalous agents in mobile agent system: a
  preliminary approach,'' in \emph{Proceedings of the first international joint
  conference on Autonomous agents and multiagent systems: part 2}, ser. AAMAS
  '02.\hskip 1em plus 0.5em minus 0.4em\relax New York, NY, USA: ACM, 2002, pp.
  655--656. [Online]. Available: \url{http://doi.acm.org/10.1145/544862.544894}
\BIBentrySTDinterwordspacing

\end{thebibliography}
\end{document}